\documentclass[journal, 10pt]{IEEEtran}
\usepackage{cite}
\usepackage{soul}
\usepackage[pdftex]{graphicx}
\usepackage{graphicx}
\usepackage{subfigure}
\usepackage{amsmath}
\usepackage{mathptmx}
\usepackage{amsfonts}
\usepackage{amssymb}
\usepackage{color}
\usepackage{float}
\usepackage{color}

\newcommand\Mark[1]{\textsuperscript#1}

\begin{document}
\title{Energy bandpass filtering in superlattice phase change memories}
\author{Jyotsna Bahl\Mark{1}, Pankaj Priyadarshi\Mark{2} and Bhaskaran Muralidharan\Mark{2}
\thanks{\Mark{1}Centre for Research in Nanotechnology and Science, Indian Institute of Technology Bombay, Powai, Mumbai-400076, India}\thanks{\Mark{2}Department of Electrical Engineering, Indian Institute of Technology Bombay, Powai, Mumbai-400076, India.}}
\maketitle


\begin{abstract}
We propose energy bandpass filtering employed using the idea of anti-reflection heterostructures as a means to reduce the energy requirements of a superlattice phase change memory based on GeTe and Sb$_{2}$Te$_{3}$ heterostructures. Different configurations of GeTe/Sb$_{2}$Te$_{3}$ superlattices are studied using the non-equilibrium Green's function approach. Our electronic transport simulations calculate the coupling parameter for the high resistance covalent state, to $97 \%$ that of the stable low resistance resonant state, maintaining the ON/OFF ratio of 100 for a reliable read operation. By examining various configurations of the superlattice structures we conclude that the inclusion of anti-reflection units on both sides of the superlattice increases the overall ON/OFF ratio by an order of magnitude which can further help in scaling down of the memory device. It is also observed that the device with such anti-reflection units exhibits 32$\%$ lesser RESET voltage than the most common PCM superlattice configurations and 27$\%$ in the presence of elastic dephasing. Moreover, we also find that the ON/OFF ratios in these devices are also resilient to the variations in the periodicity of the superlattice. 
\end{abstract}


\IEEEpeerreviewmaketitle

\section{Introduction}

\IEEEPARstart {W}ith the ever increasing demand of data density, the quest for low power memories becomes inevitable. Phase change memories (PCM) based on chalcogenide materials (such as Ge$_{2}$Sb$_{2}$Te$_{5}$) are potential candidates for non-volatile random access memories, which include switching between a highly resistive amorphous phase (RESET) to a low resistance crystalline phase (SET) for their inherent operation \cite{raoux,bipin,Wong,Burr2016}. This switching operation is characterized by a unique energy-time profile involving very high currents in general, a major part of which is not utilized for switching. In order to reduce the energy requirement for switching, various design strategies have been proposed in the literature involving geometry based \cite{Pillar,Pore,Cho,Bridge,Mu_Trench,BEC_via} or bottom electrode material based approaches \cite{SiGe_Lee_2006,Hubert,Wu}. Apart from these, thermoelectric effects originating due to high local temperatures combined with high current density in the device have also been harnessed, which have been demonstrated to reduce the programming currents down to ~$100\,\mu$A \cite{Bahl2015}. Despite all these approaches, the total loss of energy in a PCM cell amounts to a huge fraction of the total energy consumed.\\
\begin{figure}
		\includegraphics[width=3.6in,height=3.8in]{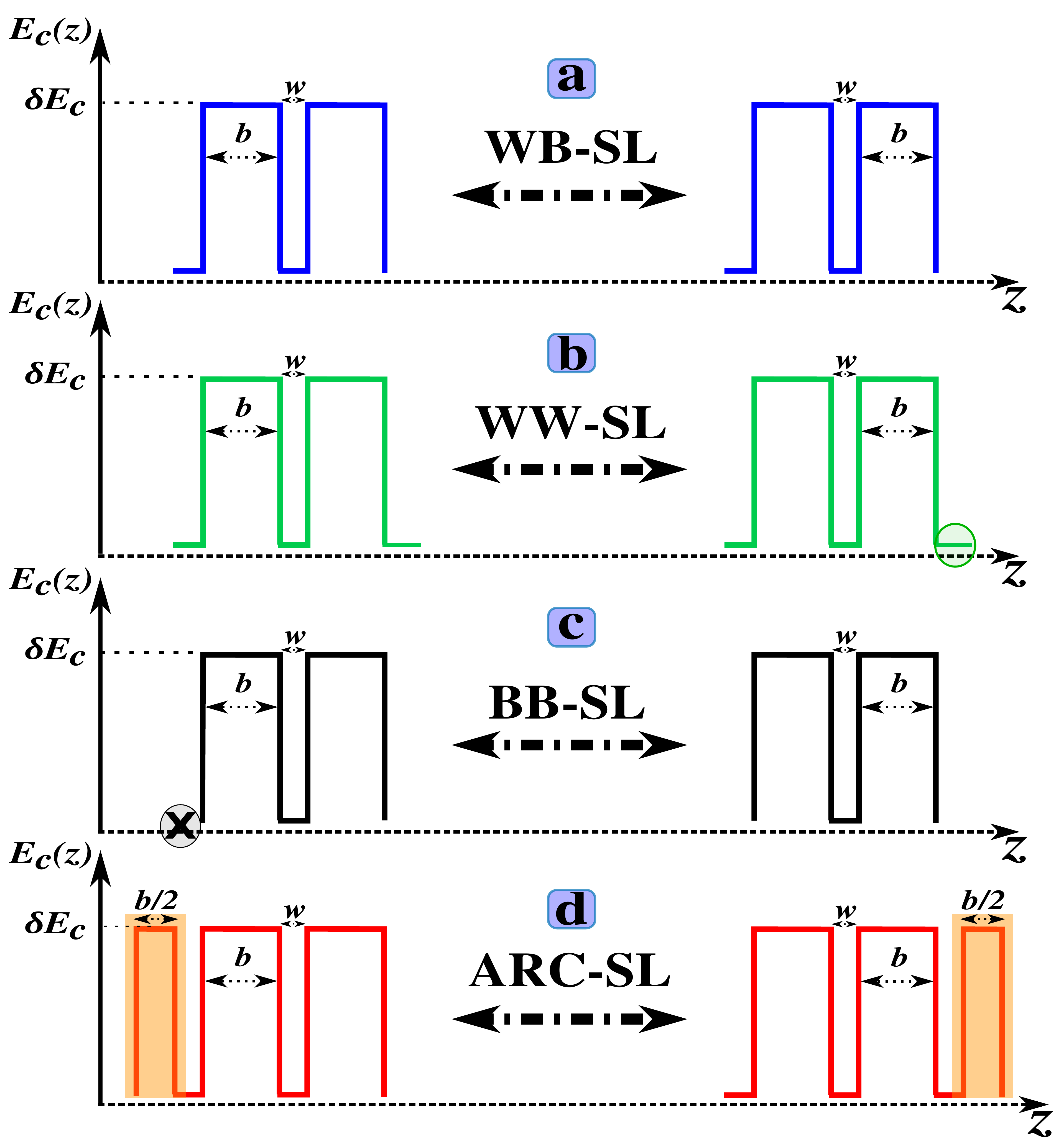}
		\caption{Device schematics: A schematic band diagram exhibiting the PCM superlattice structures. Here W denotes the 1nm GeTe Well and B denotes the 4nm Sb$_{2}$Te$_{3}$ barrier. a) A conventional iPCM device named as WB-SL configuration which has 8 periods (GeTe/Sb$_{2}$Te$_{3}$)$_{8}$ with GeTe well on the left and Sb$_{2}$Te$_{3}$ barrier on the right. b) A WW-SL configuration with an additional well (encircled with green) to the right. c) A BB-SL configuration with removal of the well (black cross showing removal of the well) from the left keeping the same number of barriers. d) The proposed ARC-SL device with two anti reflection coating barriers of half barrier width on both sides of the device (shaded region) to act as energy bandpass filters.
 }
\label{PCM_SL}
\end{figure}
\indent Another class of PCM involving superlattice-like structures (SLL) and crystalline amorphous superlattice structures (CASL) was proposed by Chong et. al., \cite{Chong2006, Chong2008} which comprise thick layers of GeTe and Sb$_{2}$Te$_{3}$ placed alternatively. Such structures involve both polycrystalline and amorphous phases unlike the single crystalline structure in a typical superlattice, and depend upon the reduced thermal conductivity of a GeTe/Sb$_{2}$Te$_{3}$ based device than that with Ge$_{2}$Sb$_{2}$Te$_{5}$ for their improved performance. However, these structures inherently involve two independent crystallization events for GeTe and Sb$_{2}$Te$_{3}$ and are limited by entropy based losses. In order to reduce the energy losses associated with thermal phase change process, Simpson et. al.\cite{Simpson2011}, proposed a GeTe/Sb$_{2}$Te$_{3}$ superlattice based interfacial PCM (iPCM) (Figure~\ref{PCM_SL}(a)) where the phase change phenomenon occurs predominantly at the interface of an ultrathin ($<2nm$) GeTe and Sb$_{2}$Te$_{3}$. \\ 
\indent Unlike the conventional Ge$_{2}$Sb$_{2}$Te$_{5}$ or SLL PCM where phase change occurs between amorphous and crystalline phases of the material involving a molten state of the material, in an iPCM, the phase change happens between two crystalline phases without the intervening melt-quench process. Several mechanisms have been proposed to explain the phase change in a superlattice PCM including polarization dependent optical control, magnetic field, electric field, thermal activation and charge injection, where charge injection is the most widely accepted \cite{Yu2015}. The charge injection mechanism \cite{Kato2013,Takaura2013} of the phase change in the iPCM has been attributed to a structural transition between the 6-fold resonantly bonded cubic structure to a 4-fold covalently bonded diamond structure caused by a short range displacement of Ge atoms in GeTe, which in itself is a unique compound with abundance of thermoelectric, ferroelectric and phase change properties despite the simple stoichiometry \cite{Boschker2017}. The work by Simpson et. al. \cite{Simpson2011} has opened up new avenues of research in the domain of PCMs to reduce the energy requirements and enhance data densities. \\
\indent This work is guided by the fact that with extremely small dimensions and elimination of amorphous phase altogether, the conducting behavior of iPCM can be controlled by principles of band engineering. Here, we propose the use of anti reflective regions as energy bandpass filters to improve the performance of superlattice PCM. For instance, anti reflective coatings (ARC) \cite{Pacher2001,Morozov,Morozov2002} have already been proposed as excellent bandpass filters to enhance the efficiency and output power of superlattice thermoelectrics \cite{Mukherjee2018,Sharma2018,Priyadarshi2018}. However, their application for programming energy reduction in PCMs have not been explored as of yet. We demonstrate quantitatively using the atomistic non-equilibrium Green's function (NEGF) simulations on superlattice PCM structures that the inclusion of ARC in conventional iPCM structures outperforms it in terms of programming energy requirements. \\
\indent This paper is organized as follows. In section II we describe the simulation methodology used in this work and in Section III we explain the key findings. Our electronic transport simulations calculate the coupling parameter for the high resistance covalent state, to $97 \%$ that of the stable low resistance resonant state, maintaining the ON/OFF ratio of 100 for a reliable read operation. By examining various configurations of the superlattice structures we conclude that the inclusion of anti-reflection units on both sides of the superlattice increases the overall ON/OFF ratio by an order of magnitude which can further help in scaling down of the memory device technology. It is also observed that the device with such anti-reflection units exhibits 32$\%$ lesser RESET voltage than the most common PCM superlattice configurations. Further, we examine the impact of elastic dephasing on the device performance, which maintains the advantage of using anti-reflection units. Moreover, we also find that the ON/OFF ratio in these devices are also resilient to the variations in the periodicity of the superlattice. In Section IV, we conclude with a brief discussion. 
\section{Simulation Methodology}
\subsection{Device schematics}
Figure~\ref{PCM_SL}(a) shows a schematic one-dimensional (1D) band diagram of the conventional iPCM structure \cite{Simpson2011} with eight periodic layers of a 1nm GeTe acting as a well (W), and a 4nm Sb$_{2}$Te$_{3}$ acting as a barrier (B) with a conduction band edge offset $\Delta E_{c}$ of 0.65eV \cite{Stiles1968,Yavorsky2011,Chua2011,Hu2011}, thus forming a 40nm wide superlattice channel. For conduction to take place, there are source and drain electrodes to the left and the right of superlattice channel (not shown in figure). This configuration is labeled as WB-SL configuration since the left and right layers (also viewed as top and bottom layers) of superlattice are comprised of well (W) and barrier (B) materials respectively. It is to be noted that as per the superlattice nomenclature "Well" refers to the layer sandwiched between two barriers of different materials. However, here we refer to the layer with \textit{well material} as W unless otherwise mentioned. On application of an electrochemical potential gradient via the contacts along the $\hat{z}$ direction, conduction takes place through the superlattice and a critical potential leads to the phase change in the device, transforming the low resistance state (LRS) to a high resistance state (HRS) and vice versa. \\
\indent As will be explained in subsequent sections, the ARC consists of two thin (half width of superlattice barrier) layers of barrier material on both sides of the superlattice which results in an improved conduction of the device. Therefore, for incorporating ARC in WB-SL configuration, either we need to add an additional well to the right (encircled with green) in WB-SL configuration to make it WW-SL configuration (Figure~\ref{PCM_SL}(b)) or remove a well (black cross showing removal of the well) from the left to make it BB-SL configuration (Figure~\ref{PCM_SL}(c)), keeping the number of barriers constant. Figure~\ref{PCM_SL}(d) shows the proposed ARC-SL configuration which includes ARC coatings on both sides (shaded region) of the PCM superlattice. The barrier height is kept constant in all the configurations of figure~\ref{PCM_SL}. Furthermore, we emphasize that WW-SL structure is not favorable for fabrication \cite{Ohyanagi2014} and is only included in this work for the sake of comparison.
\subsection{Electronic transport simulations}
A quantum transport analysis of superlattice PCM in all the four configurations is carried out using a self consistent non-equilibrium Green's function (NEGF) formalism with the Poisson solver \cite{QTDatta,LNE}. To estimate electronic currents using this NEGF-Poisson approach, the prerequisite is the calculation of transmission spectrum $T(E)$, which is obtained using the following equation
\begin{equation}
T(E)=Tr[\Gamma_{1} G(E) \Gamma_{2} G^\dagger(E)],
\label{eqTM}
\end{equation}
where $Tr$ denotes the Trace of the matrix, $[G(E)]$ is the matrix representation of energy resolved 'retarded' Green's function given by
\begin{equation}
[G(E)]=[E\textbf{I}-H-U-\Sigma(E)]^{-1},
\label{eqG}
\end{equation}
where $\textbf{I}$ is the identity matrix,  $[H]$ is the effective mass Hamiltonian calculated using nearest neighbor tight binding model, $[U]$ is the total effective potential and $\Sigma(E)$ is the self energy matrix representing coupling of $[H]$ with contacts. $\Gamma_{1(2)}$ in equation (\ref{eqTM}), represents the broadening matrix of left and right contacts respectively and is given by
\begin{equation}
\Gamma_{1(2)}(E)=i[\Sigma_{1(2)}(E) - \Sigma_{1(2)}^\dagger(E)],
\label{eqGam}
\end{equation}
The potential $[U]$ in equation (\ref{eqG}) includes collaborative effect of externally applied potential and electrostatic potential in the device, and is solved self consistently using the Poisson’s equation given by
\begin{equation}
\frac{d}{dz}\Big(\epsilon_r\frac{d}{dz}U(z)\Big) = \frac{q^2}{\epsilon_0} [N_{D} - n(z)],
\label{eqPoisson}
\end{equation}
where $(\hat{z})$ is the direction of charge transport, N$_{D}$ is the doping density and $n(z)$ is the electron density given by
\begin{equation}
n(z) = \frac{1}{a_{0}} \int\frac{G^n(E)}{2 \pi} dE,
\label{enz}
\end{equation}
where $a_{0}$ is the interatomic spacing and $G^n$ is the diagnonal element of electron correlation function given by
\begin{equation}
[G^n(E)]=[G][\Gamma_{1} f_1 + \Gamma_{2} f_2][G]^{\dagger},
\label{eqGn}
\end{equation}
where $f_{1(2)}$ is the Fermi-Dirac distribution of the left (right) contact. Once transmission is calculated, it can be used to calculate currents by using the Landaur equation
\begin{equation}
I = \frac{q}{h}\int dE\ T(E)\ [f_1(E-\mu_1) - f_2(E-\mu_2))],
\label{eqI}
\end{equation}
where $\mu_{1(2)}$ is the electrochemical potential of contact 1(2). Using the calculated current and hence the resistance of the device in the SET and RESET state, we analyze the performance of a superlattice PCM device. It should be noted that all the simulations in this work involve electron-electron interactions via Poisson's solver as the focus of this work is primarily to engineer the electronic part. Later, we quantify the effect of elastic interactions on the overall performance of the device.

\subsection{Hamiltonian calculation}
\begin{figure}
	\centering
		\includegraphics[width=3.6in,height=2.0in]{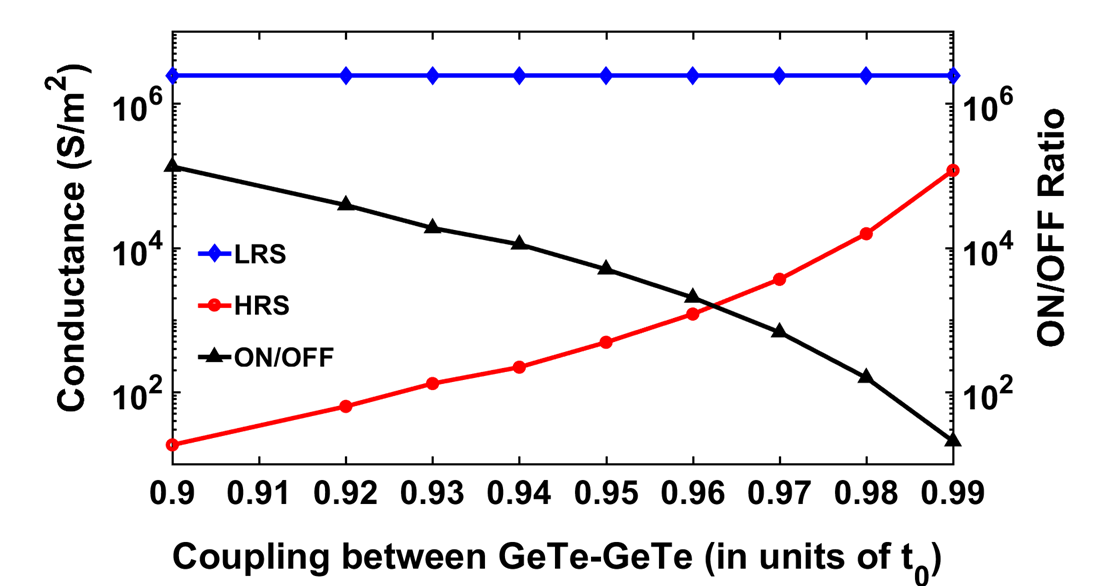}
		\caption{Calculation of coupling parameter for the Hamiltonian calculation of HRS GeTe as compared to the LRS GeTe (t$_{0}$): The variation of LRS (blue curve) and HRS (red curve) conductance (left y-axis) and device ON/OFF ratio (black curve with right y-axis) at low bias, with respect to coupling parameter of HRS in conventional device \cite{Simpson2011} which consists of GeTe well on one side and Sb$_{2}$Te$_{3}$ barrier on other side of the device with 8 periods (GeTe/Sb$_{2}$Te$_{3}$)$_{8}$. To maintain an ON/OFF ratio of 100 for a reliable switching operation, coupling parameter of 0.97t$_{0}$ between GeTe-GeTe, which is $3\%$ less than that of low resistance state, is chosen for simulations.
 }
\label{coupling_constant}
\end{figure}

The most crucial input for our simulation engine is the Hamiltonian $[H]$, which can be calculated either by using computationally expensive density functional theory (DFT) \cite{Liu2013} or using the nearest neighbor tight binding model, which under some valid assumptions provides equivalent outcomes \cite{QTDatta}. The tight binding theory can be applied directly when the material under consideration remains unaffected structurally during the device operation. However, in the case of phase change memories which involve a structural change from resonant to covalent bonding, the calculation of Hamiltonian becomes tricky. \\
\begin{figure}
	\centering
		\includegraphics[width=3.6in,height=2.5in]{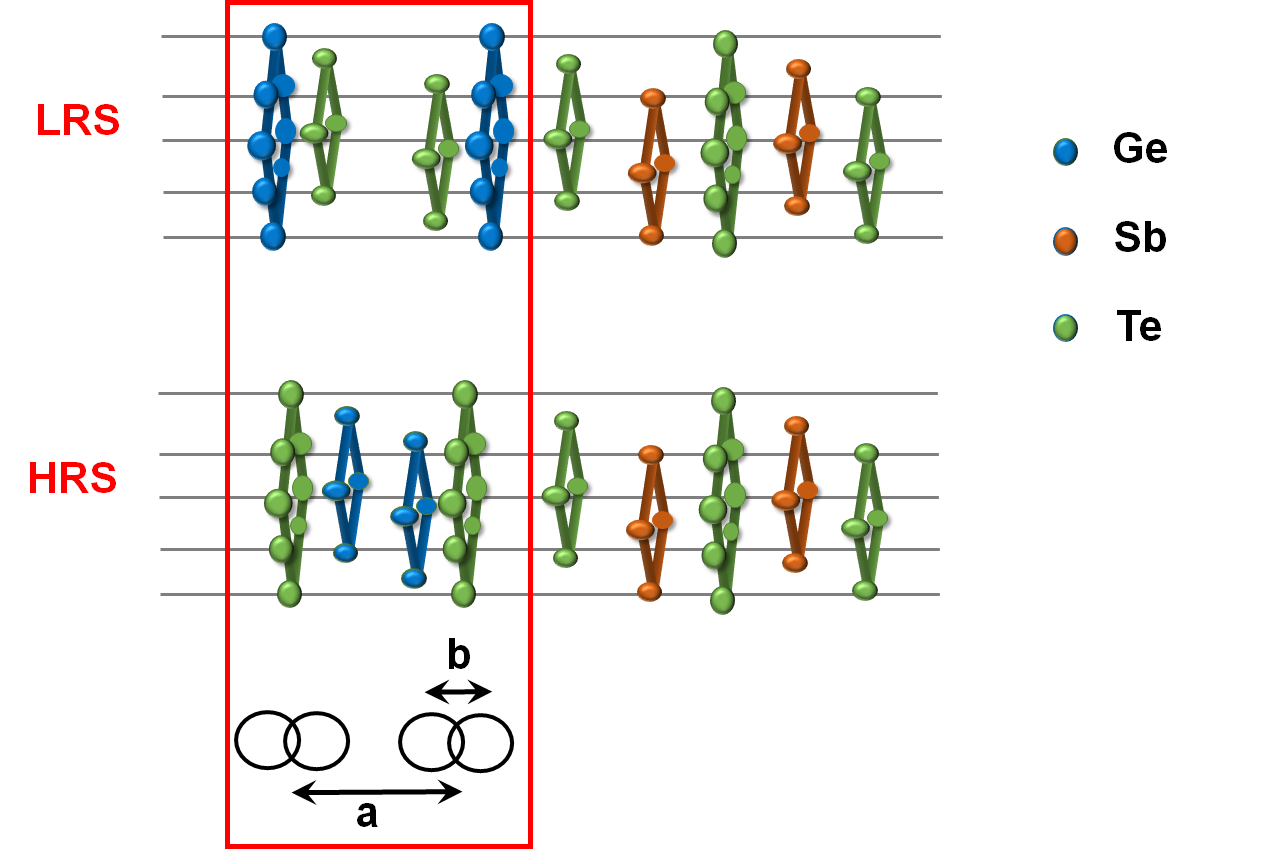}
		\caption{Switching mechanism for superlattice PCM. In the low resistance state (LRS), a 6-fold coordination exists between Ge and Te whereas in the high resistance state (HRS), Ge is 4-fold coordinated with Te, due to which there is a small local displacement of Ge atoms in the HRS with respect to the LRS. Here, it is assumed that atoms in Sb$_{2}$Te$_{3}$ remain intact in both LRS and HRS and only GeTe contributes to the phase change process. Moreover, for Hamiltonian calculations of the LRS and HRS, GeTe is considered as a diatomic molecule with different intermolecular (GeTe-GeTe) coupling constants 'a' in LRS and HRS states. Intramoleculer coupling i.e., coupling between Ge-Te and Te-Ge 'b' is assumed to be same.
 }
\label{Hamiltonian}
\end{figure}
\indent Here we present a simplified approach for the calculation of the Hamiltonian for both phases originating from the charge injection mechanism of phase change proposed in the literature \cite{Kato2013,Takaura2013}. The low resistance state (LRS) is resonantly bonded with a 6-fold coordination between Ge and Te whereas in the high resistance state (HRS), Ge is covalently bonded with Te with a 4-fold coordination, due to which there is a small local displacement of Ge atoms in HRS with respect to LRS. Figure~\ref{Hamiltonian} shows schematically the basic difference between LRS and HRS as per the charge injection mechanism. Under these assumptions, our one dimensional superlattice PCM structure with two atoms per unit cell in GeTe layer, can be treated in a manner similar to Peierls' distortion. It should be noted that while GeTe is well known to exhibit Peierls' distortion at around its Curie temperature \cite{Boschker2017}, the phenomenon still gives us cues to calculate the effective mass Hamiltonian using the nearest neighbor tight binding approach. GeTe is considered as a diatomic molecule with different intermolecule (GeTe-GeTe) coupling constants 'a' in LRS and HRS states as shown in figure~\ref{Hamiltonian}. The intramolecular coupling i.e., the coupling between Ge-Te and Te-Ge 'b' is assumed to be same. Furthermore, it is assumed that atoms in Sb$_{2}$Te$_{3}$ remain intact in both LRS and HRS and only GeTe contributes to phase change process. \\
\indent Despite GeTe and Sb$_{2}$Te$_{3}$ having a comparable conductivity effective mass (0.045m$_{0}$, where m$_{0}$ is free electron mass) \cite{Tsu1968,Chen2013,Yavorsky2011}, we have employed a spatially varying effective mass approach, which is a standard approach known to calculate the Hamiltonian of heterostructures with comprising units of different effective masses \cite{QTDatta}. For the Hamiltonian calculation of LRS, both intermolecular coupling 'a' and intramolecular coupling 'b' are considered equal to t$_{0}$, where 
\begin{equation}
t_{0W(B)} = \frac{qh^2}{16\pi^2 a^2 m_{eW(B)}}
\label{eqt0}
\end{equation}
where m$_{eW(B)}$ is the effective mass of well (barrier) material as obtained from the literature. Therefore for LRS, the Hamiltonian is the same as that of an atomic chain with a single atom per unit cell instead of two.\\
\indent  From (\ref{eqt0}), it is evident that the calculation of coupling constant for different materials depends upon the effective masses of respective materials, keeping 'a' constant. On the other hand, for the HRS of GeTe, the direct effective mass and hence the coupling constant is not known in literature. So we estimate the coupling constant of GeTe HRS for our Hamiltonian calculation by exploiting the fact that for a read operation, an ON/OFF ratio of 100 needs to be maintained for a reliable operation. We decrease the intermolecular coupling 'a' with respect to that of the LRS state considering the fact that for the HRS state, the conduction should be lesser than that with the LRS, maintaining an ON/OFF ratio of 100 at the read voltage. At the same time, the intramolecular coupling 'b' is kept constant equal to t$_{0}$. A variation factor of 1$\%$ to 10$\%$ is tried and NEGF-Poisson equations are solved self consistently to calculate low bias conductance of LRS as well as of HRS state thus obtained (Figure~\ref{coupling_constant}). As expected, the conductance of HRS state (red curve) reduces with an increase in variation factor with respect to LRS, resulting in an increasing ON/OFF ratio (black curve). However, to avoid underestimation of HRS coupling constant and hence unrealistic ON/OFF ratio, we consider the variation factor at which ON/OFF ratio reaches two orders of magnitude. Therefore, we obtain a variation factor of 3$\%$ in coupling constant of HRS with respect to coupling constant of LRS. \\
\indent With this background, we carry out the simulations of the superlattice PCM configurations depicted in figure~\ref{PCM_SL} using a two step approach, i.e.,  one with a low resistance 6-fold resonant state and the other with a 4-fold covalent state, and analyze the performance of proposed superlattice PCM cell with ARC in comparison to the conventional WB PCM cell.
\section{Results and discussion}
\subsection{Transmission Characteristics}
\begin{figure}
	\subfigure[]{\includegraphics[width=1.7in,height=1.6in]{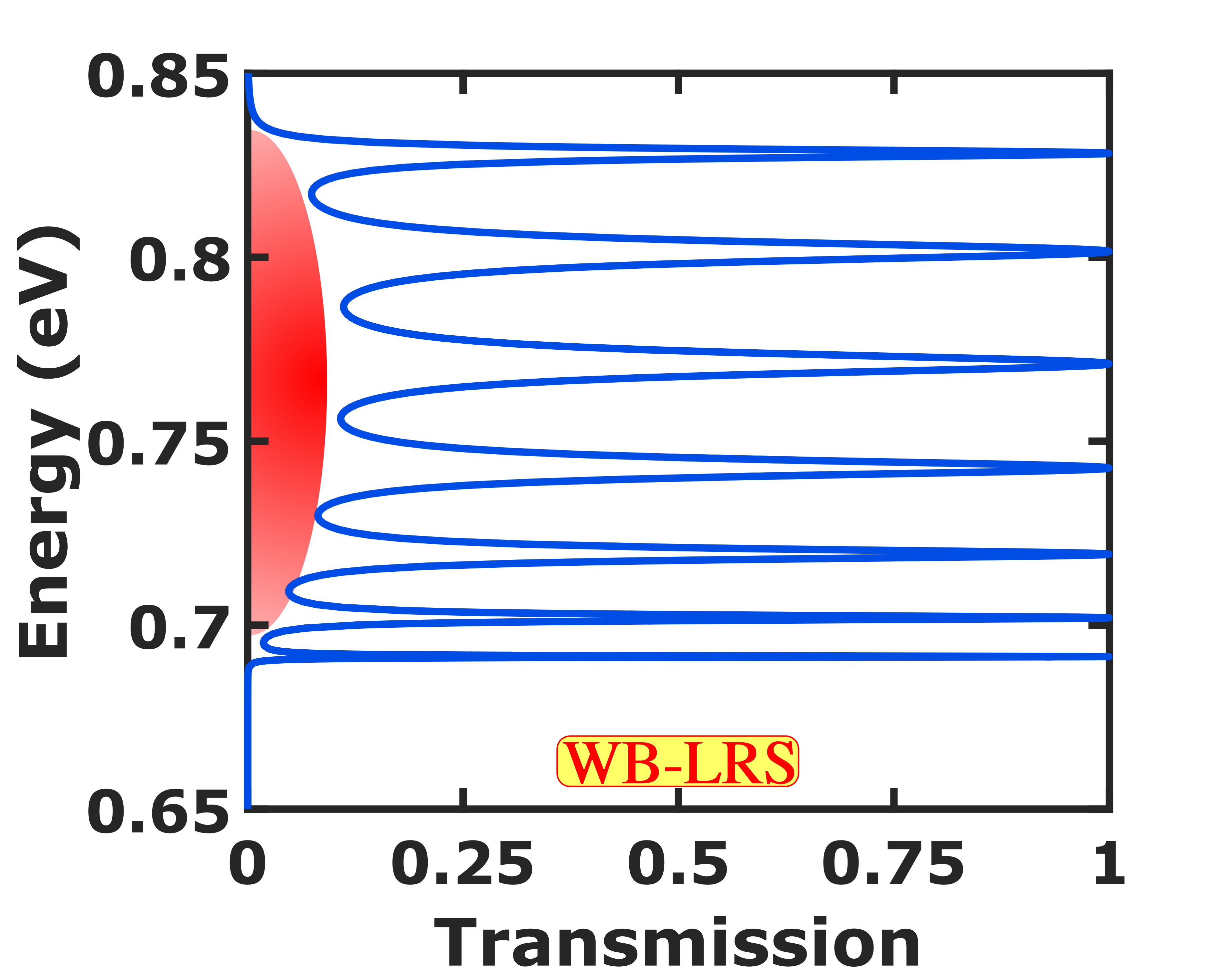}\label{WB-LRS}}
	\subfigure[]{\includegraphics[width=1.7in,height=1.6in]{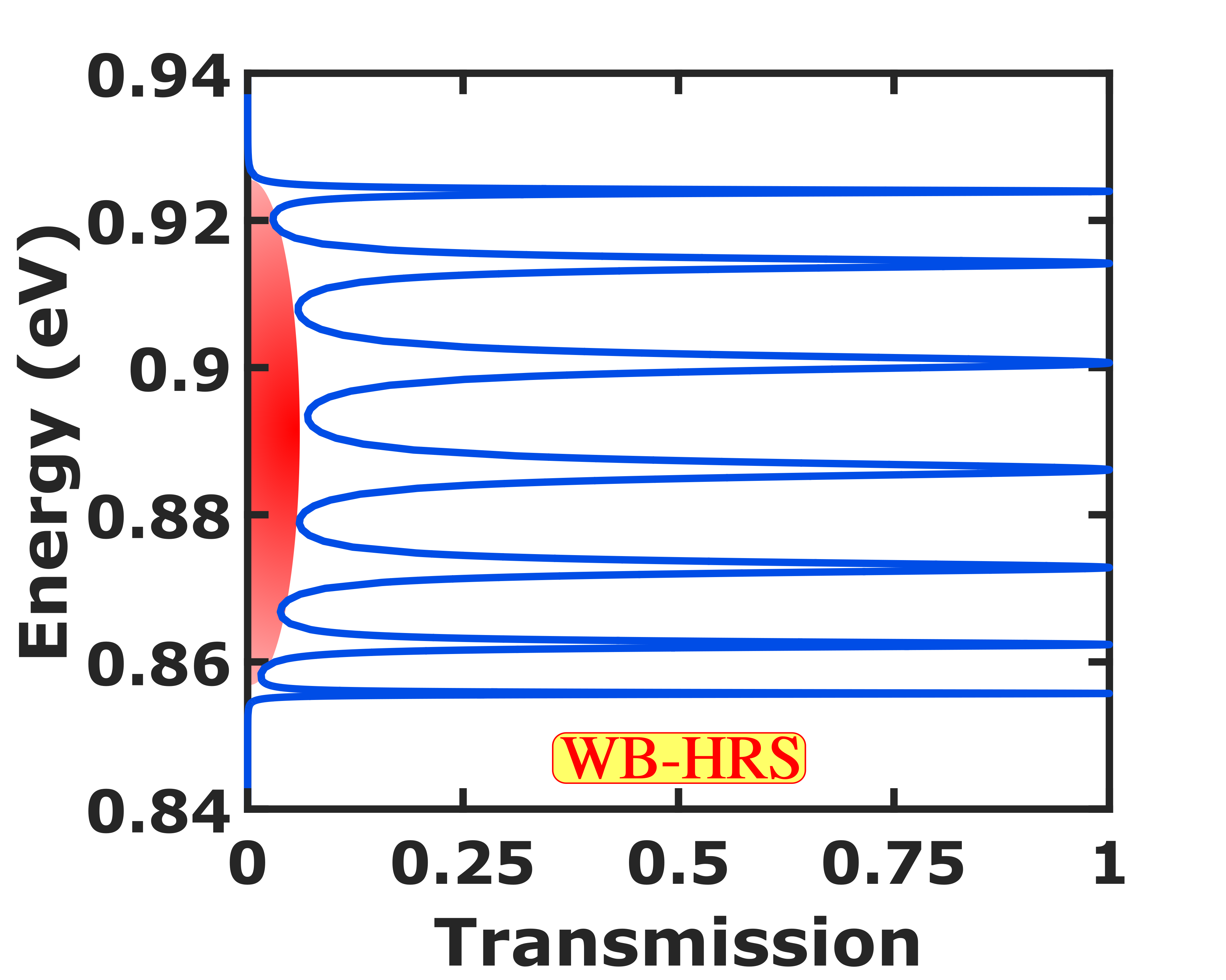}\label{WB-HRS}}
	\subfigure[]{\includegraphics[width=1.7in,height=1.6in]{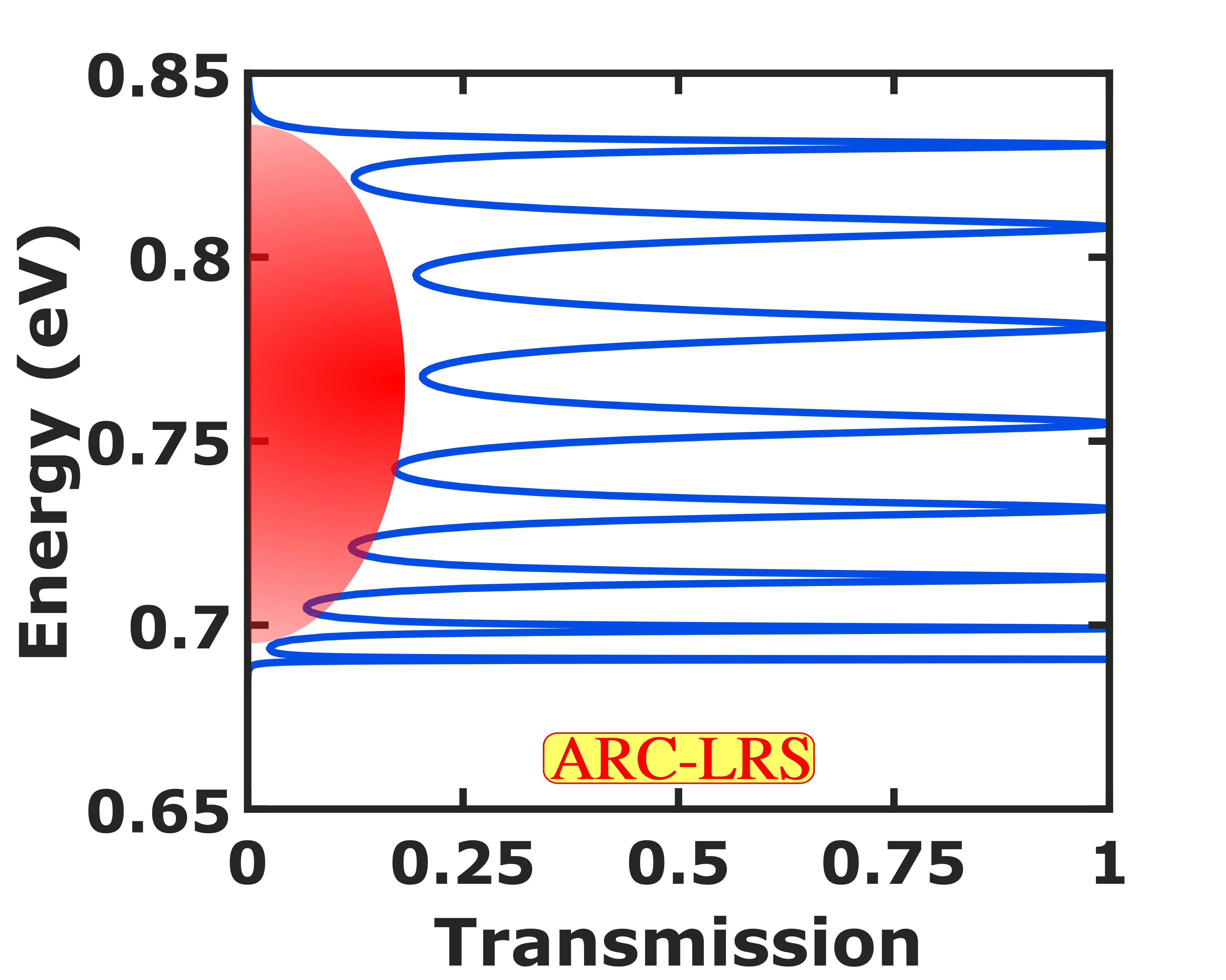}\label{ARC-LRS}}
	\subfigure[]{\includegraphics[width=1.7in,height=1.6in]{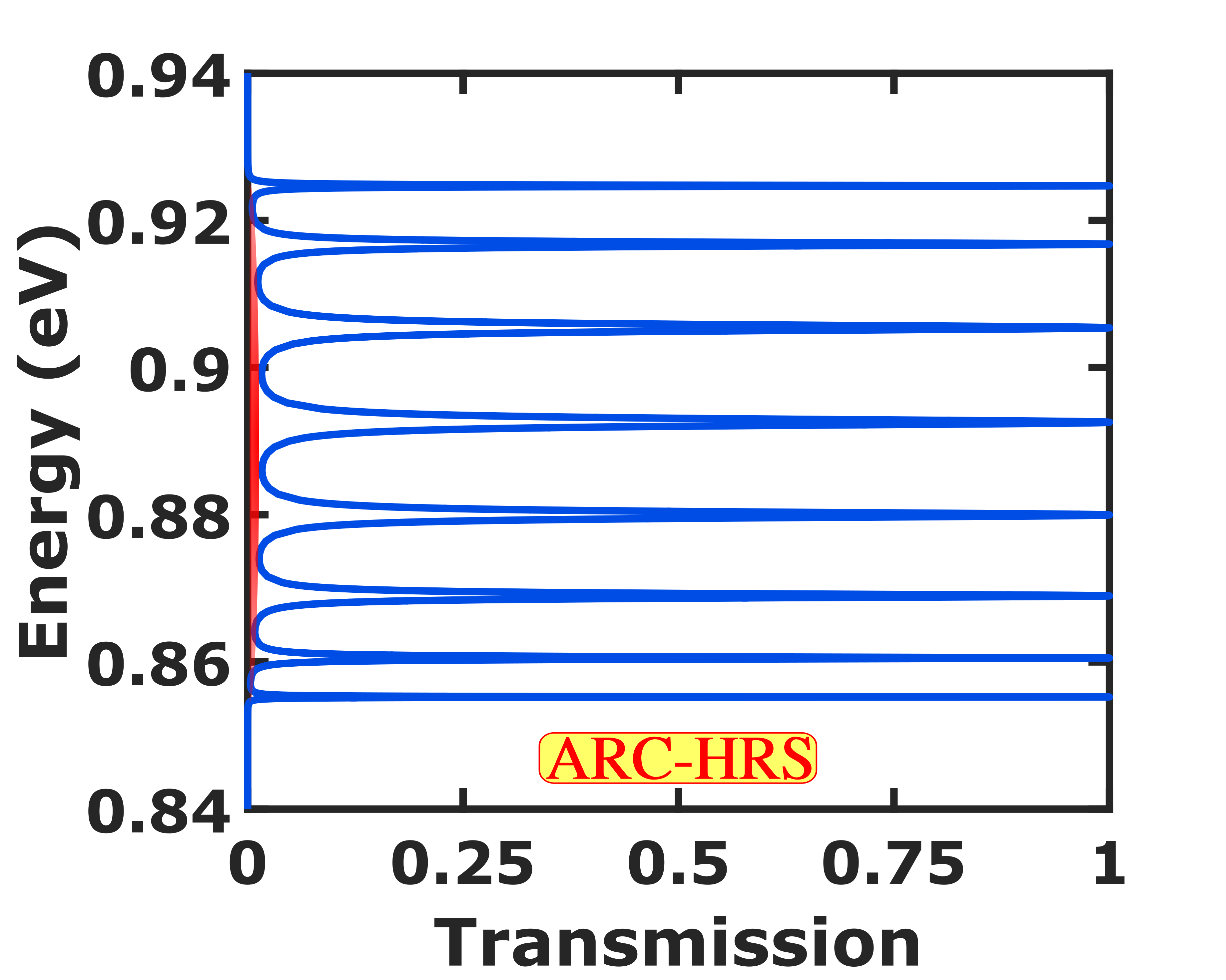}\label{ARC-HRS}}
	
	\caption{Equilibrium transmission of PCM superlattice for conventional device \cite{Simpson2011} with WB configuration i.e. GeTe well on one side and Sb$_{2}$Te$_{3}$ barrier on other side of the device with 8 periods (GeTe/Sb$_{2}$Te$_{3}$)$_{8}$, (a) in LRS state and (b) in HRS state. (c) Equilibrium transmission of PCM superlattice with 8 periods with ARC barriers on each side of the device in LRS state and (d) in HRS state. Red shaded region depicts the area under the curve (AUC) for transmission. It should be noted that the area under the transmission curve (AUC) includes the transmission peaks as well, however, for the sake of representation we have only used the red shaded regions without filling the whole area. As expected, transmission in HRS of WB configuration is lesser than LRS. Comparing (a) and (c), it is observed that the transmission in LRS state increases with inclusion of ARC, whereas for ARC HRS state, transmission decreases with ARC (d) as compared to WB-HRS which in turn favors the ON/OFF ratio. Plots are zoomed in only for the first miniband.}
	\label{Transmission_combined}
\end{figure}

Figure~\ref{Transmission_combined} shows the transmission characteristics of the LRS and the HRS states of the superlattice with ARC in comparison with the conventional WB-SL configuration. Due to the inherent periodic nature of a superlattice, the solution of Schr\"odinger equation using the self consistent NEGF equations exhibits the miniband formation with transmission spectrum having the number of peaks equal to the number of wells in between the barriers as evident from figure~\ref{Transmission_combined}. The red shaded region in the figure shows schematically the area under the transmission curve (AUC) which is an indicative of the overall conduction in the device keeping all the other parameters constant. It should be noted that the area under the transmission curve (AUC) includes the transmission peaks as well, however, for the sake of representation we have only used the red shaded regions without filling the whole area. In accordance with the phase change memory theory, the equilibrium transmission $T(E)$ in HRS state of the WB-SL configuration in figure~\ref{WB-HRS} has a lesser effective AUC as compared to that of LRS state figure~\ref{WB-LRS}. Furthermore, the energy levels E at which transmission takes place, shift to higher side in HRS state in congruence with its lower conductivity than LRS state. \\
\indent The concept of ARC is well known in optics, and is employed to reduce the reflection from the lens surfaces. On a similar note, the wave nature of electrons permits the modulation of the overall transmission through a device by minimizing the overall reflection and hence improving the transmission. As a primary condition of ARC, the additional barrier should be equal to half the barrier width of superlattice to act as a Bragg reflector and with a potential (barrier height) commensurate with the Bloch eigen-states of the periodic superlattice encompassed by the ARC. Figure~\ref{ARC-LRS} shows the equilibrium transmission of the superlattice with the ARC barriers on both sides of the superlattice. It is observed that in the LRS, the device with ARC has a better transmission in the first miniband as compared to that of the WB-SL structure. However, in the HRS, there is a reduction in the transmission as seen in figure~\ref{ARC-HRS} favoring the ON/OFF ratio of the PCM device. 

\subsection{Switching voltage}
\begin{figure}
	\centering
		\includegraphics[width=3.3in,height=2.1in]{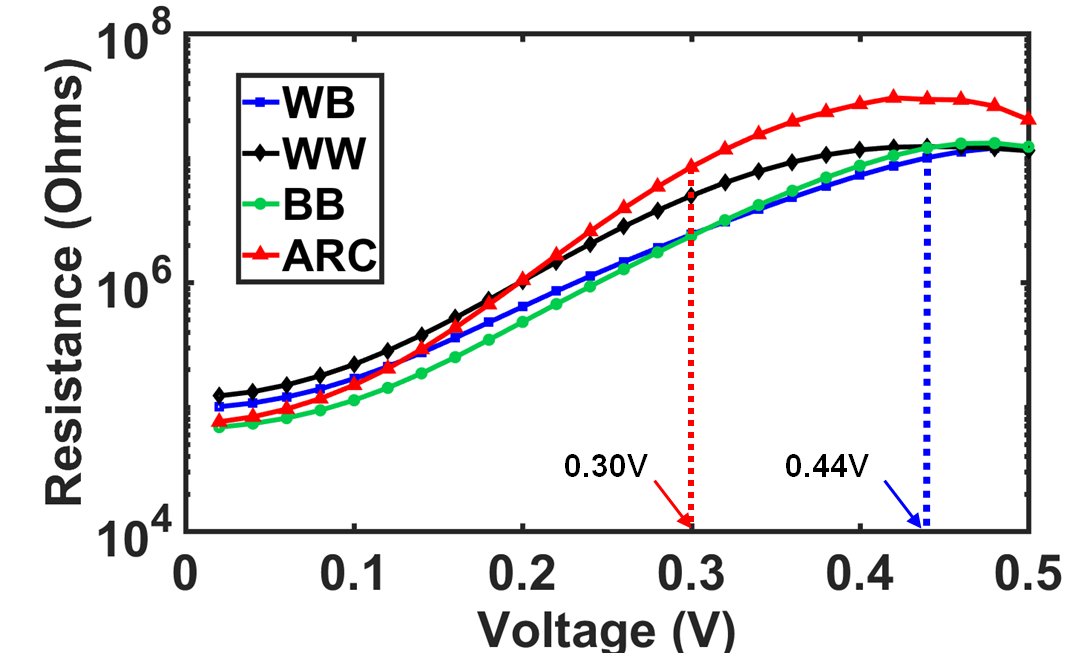}
	\caption{R$_{RESET}$ vs voltage plot for LRS state of four configurations used in the study. Cut off voltage is calculated when ON/OFF ratio becomes 100. For ARC configuration (red curve), a cutoff voltage of 0.30V is noted as compared to 0.44V with WB configuration (blue curve) which amounts to 32$\%$ reduction}
	\label{IV}
\end{figure}

As explained in previous section, the ARC helps in attaining a better transmission, which in turn reflects in the overall conduction of the device and thereby a reduction in the threshold voltage or programming voltage, as will be explained here. Figure~\ref{IV} shows the variation of R$_{RESET}$ with respect to the applied voltage. To calculate R$_{RESET}$ or R$_{SET}$, a voltage bias is applied to the device and the current is calculated using Landauer equation. It is observed that the R$_{RESET}$ increases with an increase in voltage. The threshold (or programming) voltage is calculated when the R$_{RESET}$/R$_{SET}$ or the ON/OFF ratio reaches 100 which is a reliable cutoff for storing the bits. With the inclusion of ARC, the programming voltage is noted to be 0.3V as compared to 0.44V with WB-SL configuration, amounting to a significant 32${\%}$ reduction.
\subsection{Low bias conductance}
\begin{figure}
	\subfigure[]{\includegraphics[width=3.4in,height=2.2in]{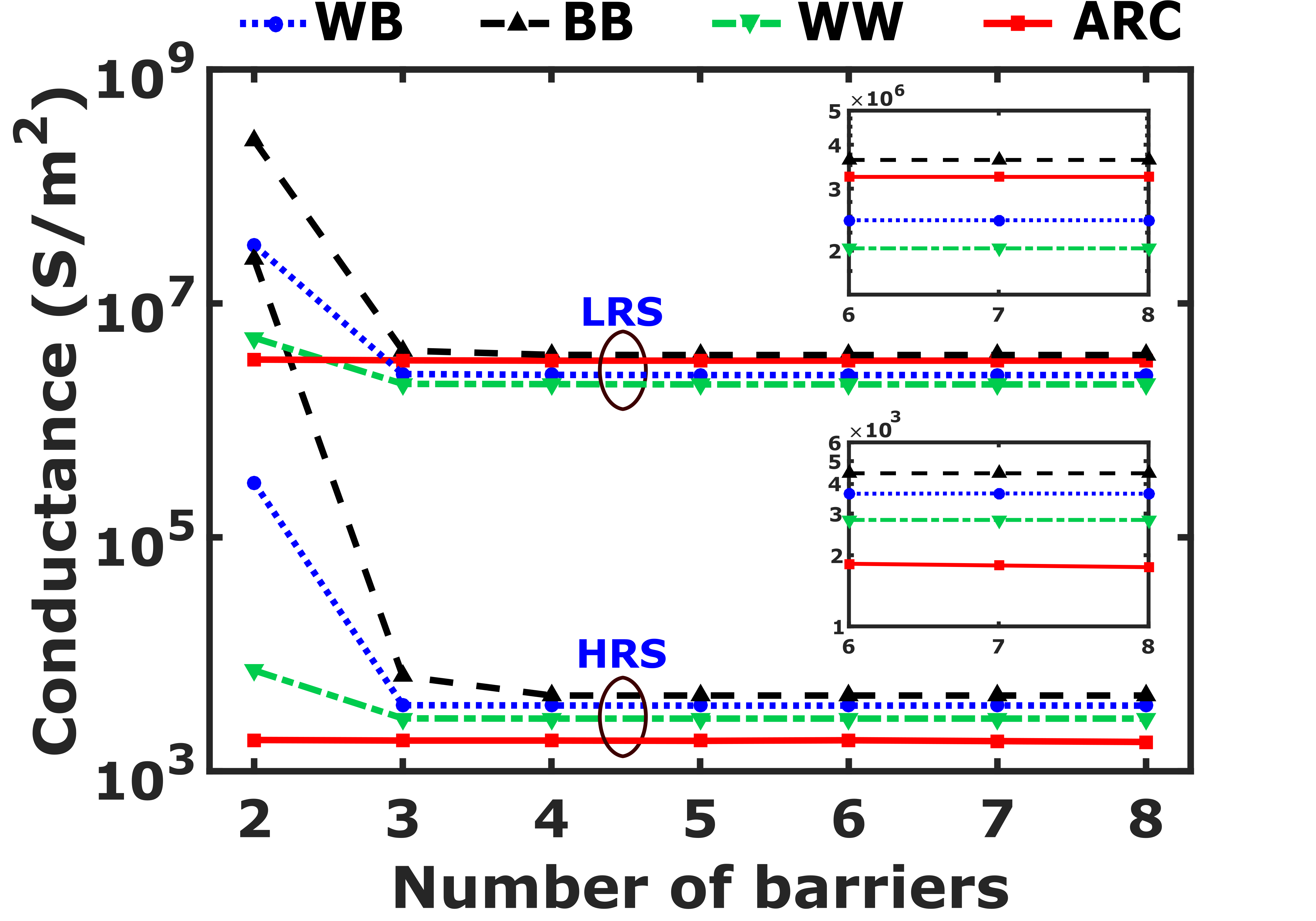}\label{Conductance_Barrier}}
	\subfigure[]{\includegraphics[width=3.4in,height=2.2in]{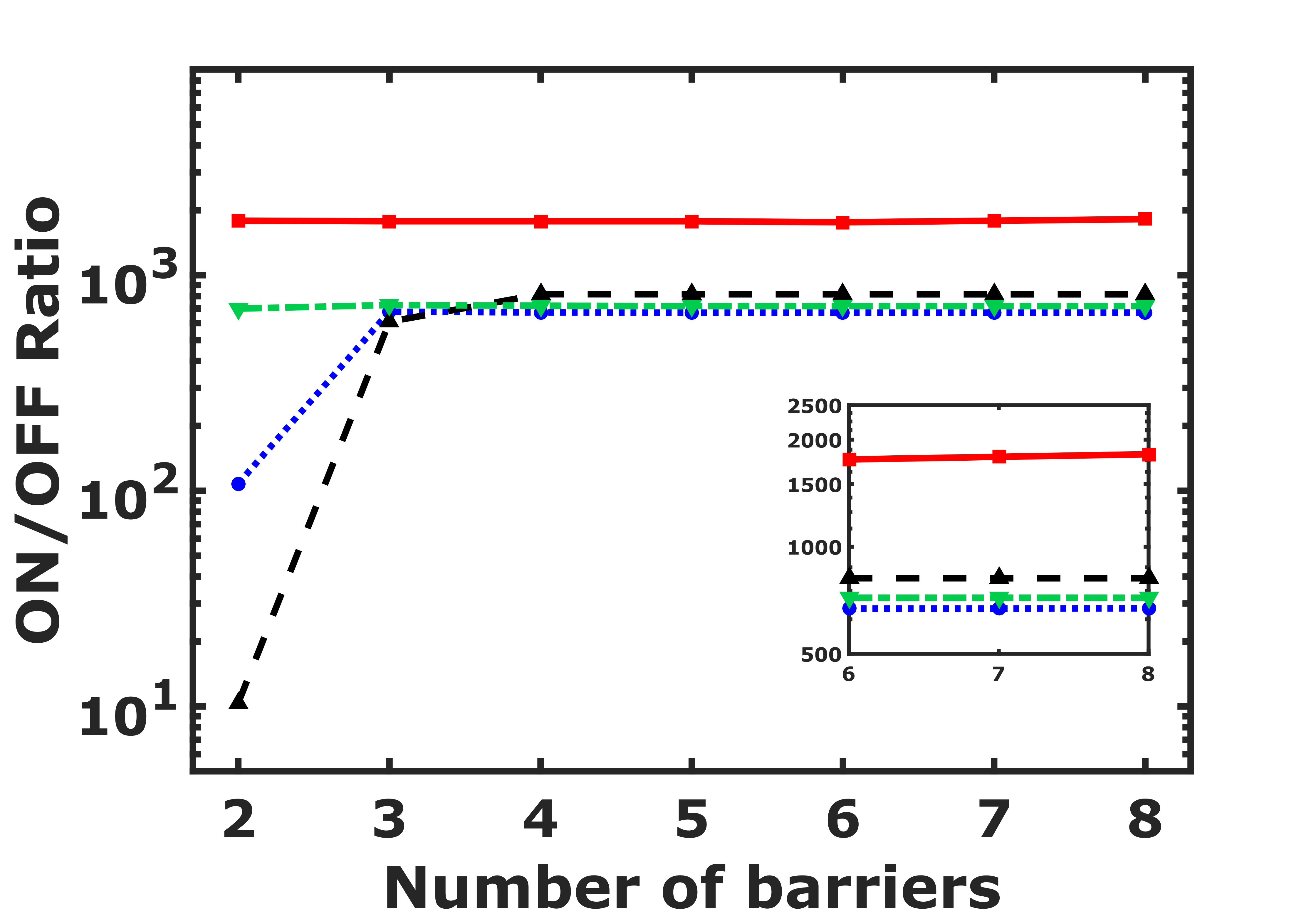}\label{Ratio}}
	\caption{(a) Low bias conductance of superlattice PCM for all the four configurations used in this work, with respect to number of barriers in the device: Solid curves show low resistance state (LRS) conductivities and dotted curves show high resistance state (HRS) conductivities. It is noted that with inclusion of ARC barriers on both sides of the device, conductance increases in low resistance state (red solid curve) and decreases in high resistance state (red dotted curve) as compared to conventional device with GeTe well on one side and Sb$_{2}$Te$_{3}$ barrier on the other side (blue solid and dotted curves respectively). (b) ON/OFF ratio of superlattice PCM for different configurations: With inclusion of ARC barriers on both sides of the device (red curve), ON/OFF ratio increases as compared to conventional device WB (blue curve).}
	\label{LBC}
\end{figure}

Figure~\ref{Conductance_Barrier} shows the conductance at a low bias (1mV), equivalent to the read out voltage of the PCM superlattice for all the four configurations used in this work, with respect to the size of the device i.e., the number of barriers. The blue dotted curve in the upper part of figure shows the low bias conductance G$_{LRS}$ in the LRS state of WB-SL PCM which remains almost constant on scaling down and increases steeply below 3 barriers i.e., 15nm which is in alignment with the literature \cite{Liu2013}. The BB-SL and WW-SL shown by the black dashed and green dot-dash curves respectively, follow the same trend as that of WB-SL. However, in the case of BB-SL, the conductance increases with respect to WB-SL and the increase is more sharp on scaling down, whereas for the WW-SL conductance in the LRS decreases as compared to the WB-SL. On the other hand, the conductance G$_{LRS}$ of ARC-SL is comparable to that of BB-SL (as shown in inset) and remains unaffected by scaling down.\\
\indent The second half (lower region) of figure~\ref{Conductance_Barrier} shows the low bias conductance of the corresponding HRS state of the devices. The scaling down behavior of the low bias conductance G$_{HRS}$ states remain similar to the LRS state, except for the sharper increase in G$_{HRS}$ on scaling down, which results in an overall decrease in the ON/OFF ratio (Figure~\ref{Ratio}) below two orders of magnitude which is critical for phase change operation, and hence puts a scaling limit on the device. However, with the ARC, the overall ON/OFF ratio improves as compared to the WB-SL which in turn will help in scaling down the device. This also explains the merit of ARC despite its conductance being lesser than that of the BB-SL in its LRS state.

\subsection{Effect of Fermi level}
\begin{figure}[ht]
	\centering
		\includegraphics[width=3.6in,height=2.4in]{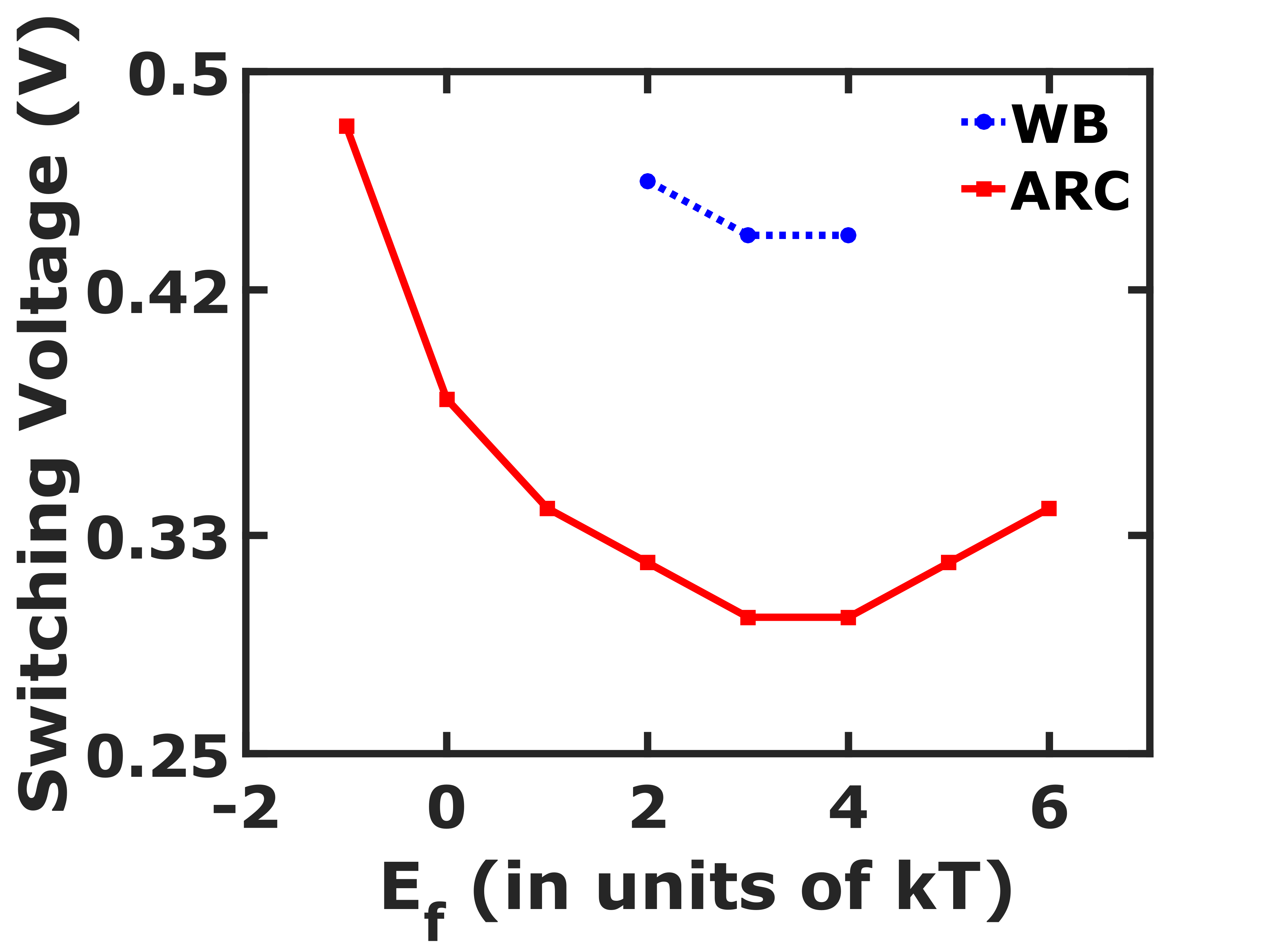}
	\caption{Variation of Switching voltage for WB and ARC devices for 8 barriers with Fermi level ($E_f$). Switching voltage is least for $E_f$=3kT and 4kT, and increases with increase or decrease in $E_f$ which is due to the relative position of Fermi level w.r.t miniband region. Moreover, for all the $E_f$ values, switching voltage is lower for ARC as compared to that of WB configuration.}
	\label{Switching}
\end{figure}

After establishing the effect of the ARC on SL-PCM performance, we shift our attention to the effect of Fermi level on switching voltage. It must be noted that for all the simulations carried out in this work, an $E_f$ of $E_c+4kT$ is assumed, which falls in the first miniband region for a reasonable conduction. Figure~\ref{Switching} depicts the importance of choosing an optimum $E_f$ for the best performance of SL-PCM. For $E_f$ values less than $2kT$ or greater than $4kT$ with respect to $E_c$, the switching voltage increases due to a mismatch of the selected energy range within the first miniband. Moreover, there is no switching outside the depicted range of $E_f$ since the required ratio for calculation of switching voltage is not achieved in case of WB-SL (blue curve). On the other hand, in the case of the ARC enabled SL-PCM, which is demonstrated to have a higher ON/OFF ratio than that with a WB-SL PCM, a broader permissible range of $E_f$ is possible which makes the device applicable for a larger range of dopings. However, the voltage required to switch the device will be larger as we move away from the optimum $E_f$ value of 3-4 $kT$ with respect to $E_c$ for the device dimensions deployed in this work.

\subsection{Effect of dephasing}
As observed in the previous sections, the inclusion of anti-reflection units in GeTe and Sb$_{2}$Te$_{3}$ superlattice PCM helps in reducing the programming energy requirements by energy bandpass filtering. The results presented up to now include only the electron-electron interactions via Poisson's solver neglecting any electron-phonon interactions. Such interactions may be phenomenologically incorporated using elastic dephasing in the device \cite{LNE,Golizadeh-Mojarad2007}. We modify equations (\ref{eqG}) and (\ref{eqGn}) to calculate self energy and inscattering functions as given below:
 \begin{equation}
\Sigma_s(E) = D_0[G(E)] 
\label{eqSigS}
\end{equation}
\begin{equation}
G(E) = [E \textbf{I}-H-U-\Sigma_1-\Sigma_2-\Sigma_S]^{-1} 
\label{eqG_new}
\end{equation}
\begin{equation}
\Sigma_{s}^{in}(E) = D_{0}[G^{n}(E)]  
\label{eqSigSin}
\end{equation}
\begin{equation}
G^{n} = [G][\Gamma_{1}f_{1}+\Gamma_{2}f_{2}+\Sigma_{s}^{in}][G^{\dagger}] 
\label{eqGn_new}
\end{equation}
where $D_{0}$ denotes scattering strength of the elastic phase-breaking event. These equations are solved self-consistently until $\Sigma_s$ and $\Sigma_s^{in}$ converge.
\begin{figure}
\centering
	\subfigure[]{\includegraphics[width=1.7in,height=1.6in]{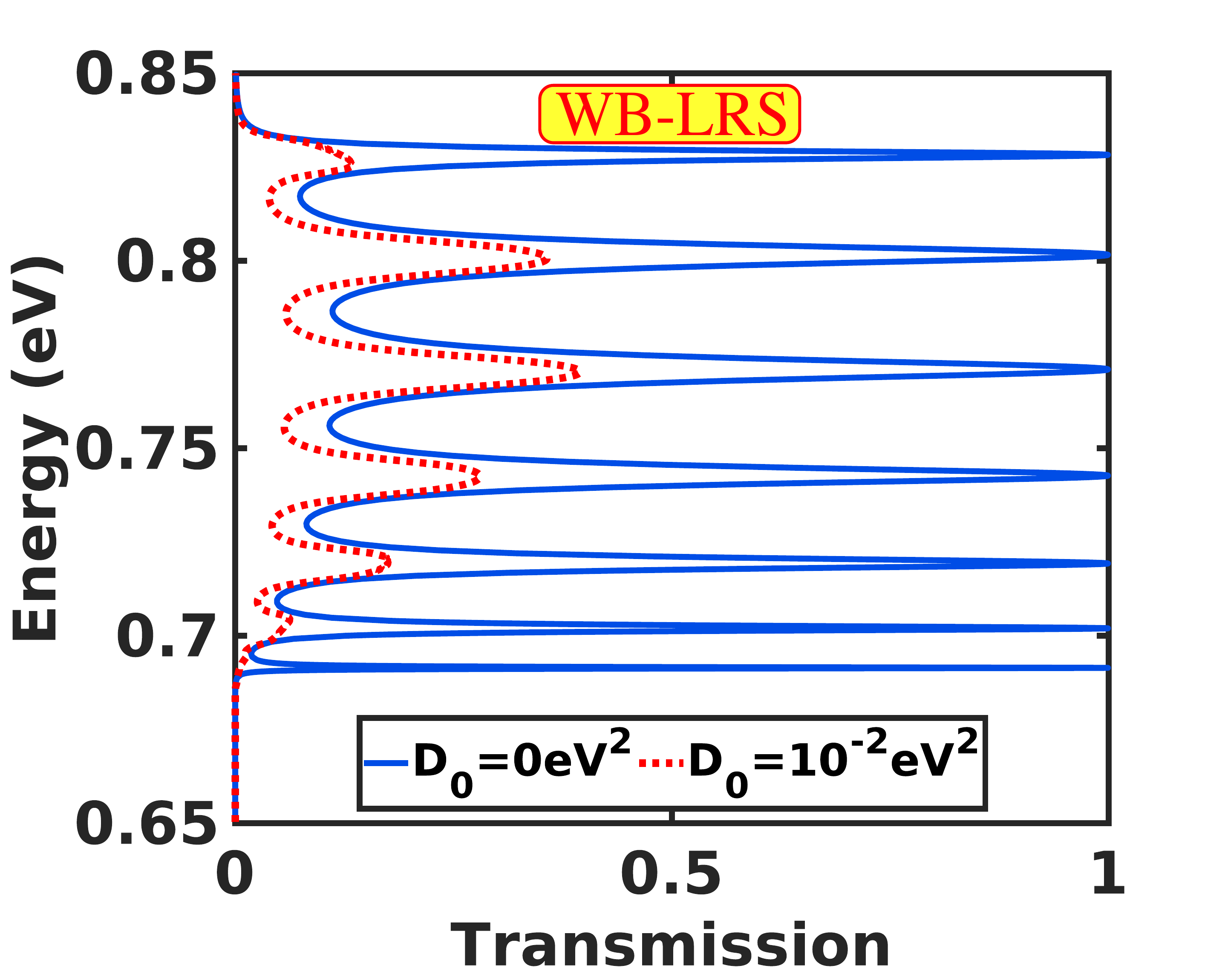}\label{WB_LRS_TM_Dephasing}}
	\subfigure[]{\includegraphics[width=1.7in,height=1.6in]{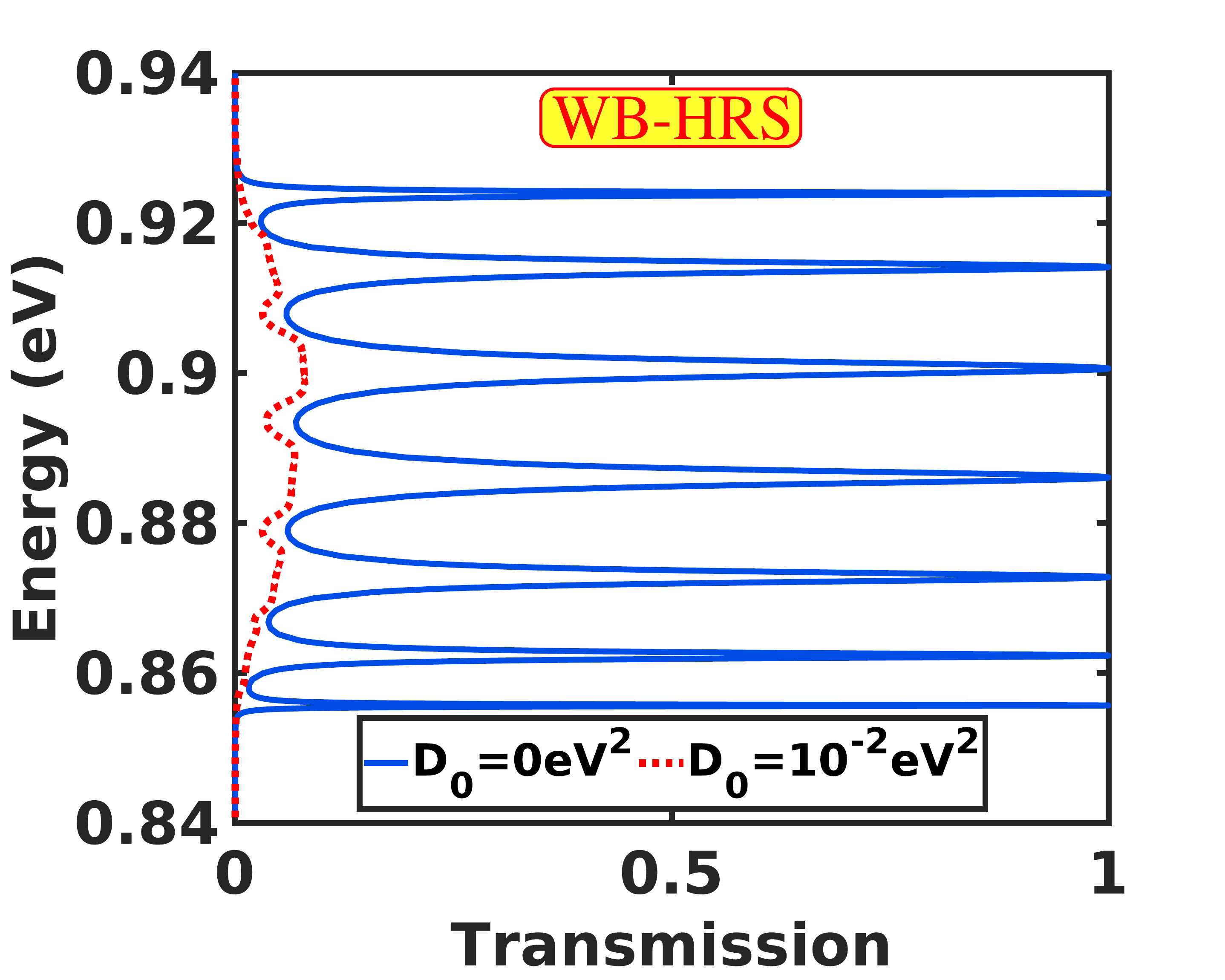}\label{WB_HRS_TM_Dephasing}}
	\subfigure[]{\includegraphics[width=1.7in,height=1.6in]{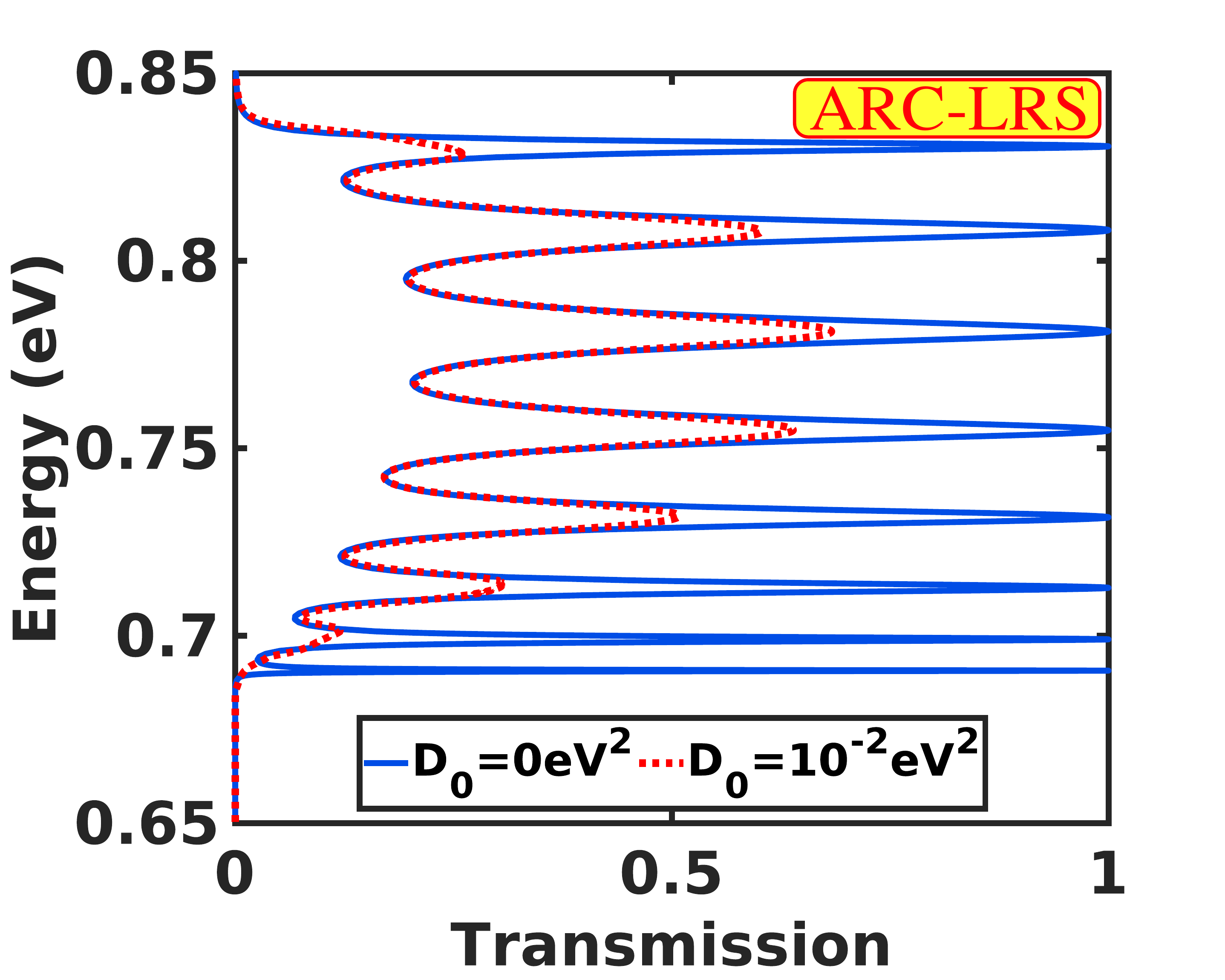}\label{ARC_LRS_TM_Dephasing}}
	\subfigure[]{\includegraphics[width=1.7in,height=1.6in]{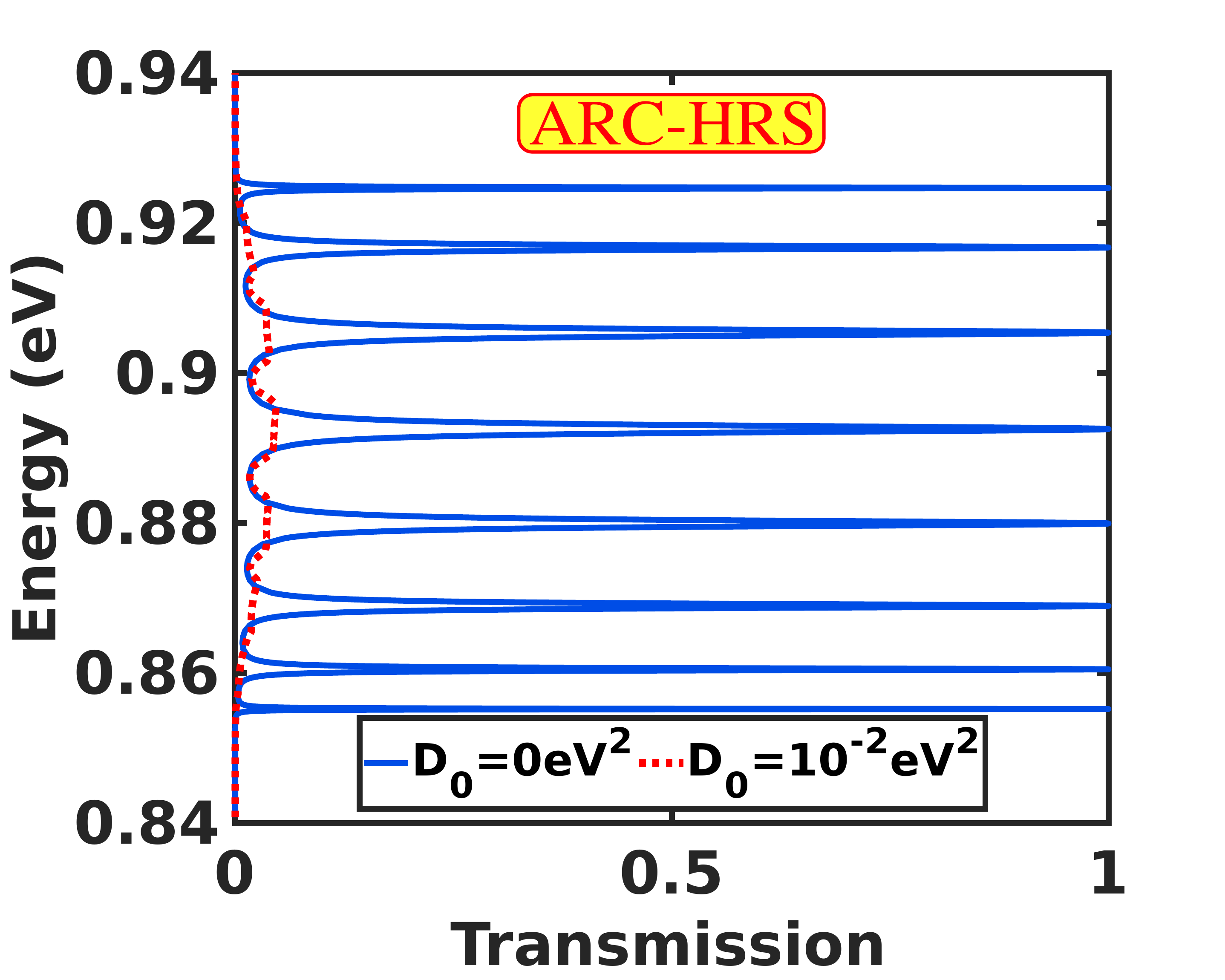}\label{ARC_HRS_TM_Dephasing}}
	\caption{Effect of elastic dephasing on equilibrium transmission of PCM superlattice for conventional device \cite{Simpson2011} with WB configuration i.e. GeTe well on one side and Sb$_{2}$Te$_{3}$ barrier on other side of the device with 8 periods (GeTe/Sb$_{2}$Te$_{3}$)$_{8}$, (a) in LRS state and (b) in HRS state. (c) Effect of dephasing on equilibrium transmission of PCM superlattice with 8 periods with ARC barriers on each side of the device in LRS state and (d) in HRS state. With inclusion of dephasing the transmission in each case decreases.}
	\label{Transmission_Dephasing}
\end{figure}

In this section, we study the effect of dephasing with a typical value of acoustic phonon scattering strength $D_{0}$ = 10$^{-2}$eV$^{2}$. The impact of dephasing on transmission curve of WB and ARC configuration in low resistance state (LRS) and high resistance state (HRS) is shown in figure~\ref{Transmission_Dephasing}.

\begin{figure}
	\centering
		\includegraphics[width=3.2in,height=2.2in]{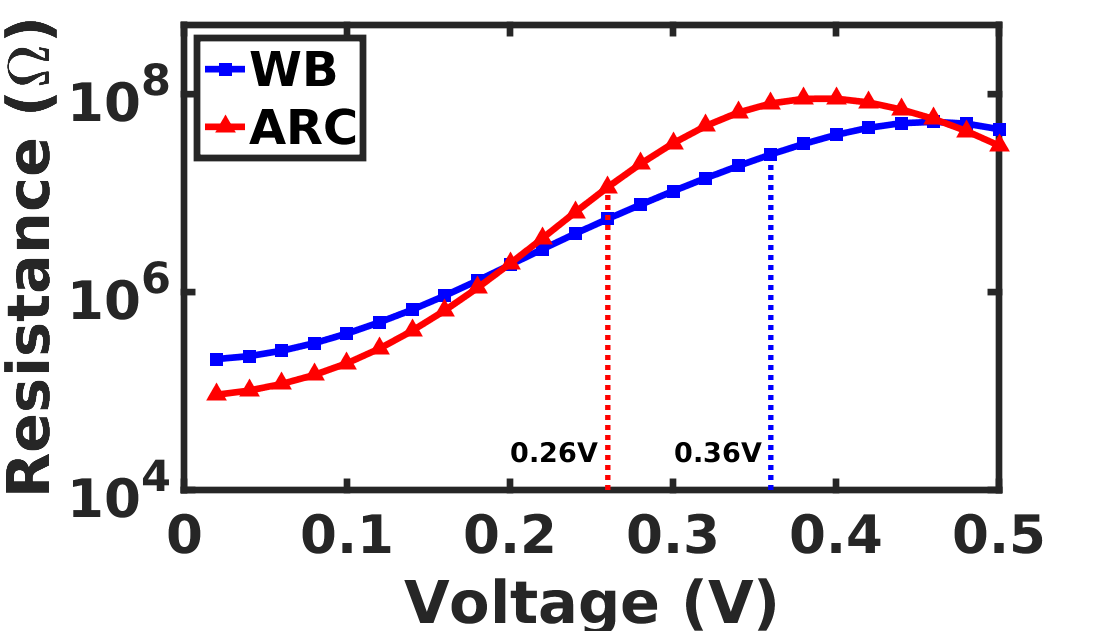}
	\caption{Effect of dephasing on R$_{RESET}$ vs voltage plot for LRS state of WB and ARC configurations used in the study. Cut off voltage is calculated when ON/OFF ratio becomes 100. For ARC configuration (red curve), a cutoff voltage of 0.26V is noted as compared to 0.36V with WB configuration (blue curve) which amounts to 27$\%$ reduction in presence of dephasing as compared to 32$\%$ without dephasing.}
	\label{RV_Dephasing}
\end{figure}

With elastic dephasing the transmission in both LRS and HRS states of WB and ARC configuration decreases. Figure~\ref{RV_Dephasing} shows the impact of dephasing on programming voltage with inclusion of ARC. It is observed that with dephasing, both for WB and ARC configurations the programming voltage decreases since it depends upon the ratio of resistances in ON and OFF states. However, the net effect of using ARC over WB configuration decreases from 32$\%$ to 27$\%$, which still emphasizes the inclusion of ARC.

\section{Conclusions}
In conclusion, we have theoretically proposed and demonstrated the use of anti reflective coating (ARC) barriers in a GeTe/Sb$_{2}$Te$_{3}$ based superlattice PCM in order to reduce the programming energy requirements of the device. Various configurations are analyzed using our self consistent NEGF-Poisson solver that was developed for this work, primarily to predict the effect of ARC on the RESET operation of the SL-PCM. It is shown that the ARC enabled superlattice PCM outperforms the conventional WB-SL PCM in terms of programming energy requirements, ON/OFF ratios and hence its scalability. The merit of using ARC over WB-SL PCM remains intact even in the presence of elastic dephasing. We believe that this work would set a stage for designing superlattice phase change memories using a physics guided approach. Furthermore, the role of electron-phonon scattering effects on the performance of ARC enabled SL-PCMs can be explored as a future problem.
\section*{Acknowledgment}
This work was supported in part by the IIT Bombay SEED grant. We also acknowledge funding from the Visvesvaraya
PhD Scheme and the Young faculty research fellowship scheme of Ministry of Electronics and Information Technology,
Government of India, being implemented by Digital India Corporation
(formerly Media Lab Asia).




\bibliographystyle{IEEEtran}
\bibliography{ref_ARCPCM}
%
%
%

%

%
%
%

\end{document}